\documentclass[superscriptaddress,article,onecolumn,preprint,showpacs,amsmath,nofootinbib]{revtex4-1}

\usepackage[english]{babel}
\usepackage{graphicx}
\usepackage{nicefrac}
\usepackage{color}
\usepackage{verbatim}
\usepackage{tikz}
\usepackage{comment}
\usepackage{setspace}

\newcommand{\beq} {\begin{equation}}
\newcommand{\eeq} {\end{equation}}
\begin{document}

\title{Random bits, true and unbiased, from atmospheric turbulence}
\author{Davide G. Marangon}
\author{ Giuseppe Vallone}
\author{ Paolo Villoresi}
\affiliation{Department of Information Engineering, University of Padova, via Gradenigo 6/B}

\date{\today}

\begin{abstract}
Random numbers represent a fundamental ingredient for numerical simulation, games, information science and secure communication.
Algorithmic and deterministic generators are affected by insufficient information entropy. On the other hand,
suitable physical processes manifest intrinsic unpredictability that may be exploited for  generating
genuine random numbers with an entropy reaching the ideal limit.
In this work, we present a method to extract genuine random bits by using the atmospheric turbulence: by sending a laser beam along
a $143 \, km$ free-space link,
we took advantage of the chaotic behavior of air refractive index in the optical propagation.
Random numbers are then obtained by converting in digital units the aberrations and distortions of the received laser wave-front.
The generated numbers, obtained without any post-processing, pass the most selective randomness tests.
The core of our extracting algorithm can be easily generalized for other physical processes.
\end{abstract}

\maketitle

It is well established that genuine and secure randomness can not be achieved with deterministic algorithms. The quest for true randomness leads to rely on physical sources of entropy, to be used in the so-called \textit{true random number generators} (TRNG).

The working principle of a TRNG consists of sampling  a natural random process and then to output an uniformly distributed random variable (namely a bit). Typical sources of entropy are the amplification of electronic noise \cite{Petrie2000}, phase noise of semiconductor lasers \cite{Reidler2009}, unstable free running oscillators \cite{Sunar2007} and chaotic maps \cite{Stojanovski2001}. In addition, a specific class of TRNG employs the intrinsic randomness of quantum processes such as the detection statistics of single photons \cite{Furst2010, stipcevic2012quantum}, entangled photons \cite{munr06pra, Pironio2010} or the fluctuations of vacuum amplitudes  \cite{Gabriel2010}. There are at least two issues with TRNGs. The first one, theoretical, deals with the fact that a chaotic physical system has a deterministic, at least in principle, evolution in time, so much study \cite{schindler2009evaluation} must be devoted in selecting those initial conditions which won't lead the system to some periodical, completely predictable trajectory \cite{DICHTL2003}. This selection can be performed by using a {robust} statistical model of the used physical system. The second problem deals with the unavoidable hardware non-idealities which spoil the entropy of the source, i.e. temperature drifts modify the thresholds levels, or the amplifier stages of photon detector make classical noise to leak inside a quantum random signal. Most of the TRNGs are then forced to include a final post-processing stage with the purpose of increasing the entropy of the emitted bits\footnote{This kind of problem involve also QRNG, which although being theoretically shielded by the postulates of Quantum Mechanics, have to deal with classical imperfect hardware. Recent literature has shown an even growing interest in developing efficient post-processing techniques to be employed in QRNG.}.

\begin{figure*}[!t]
\centering
\includegraphics[width=\textwidth]{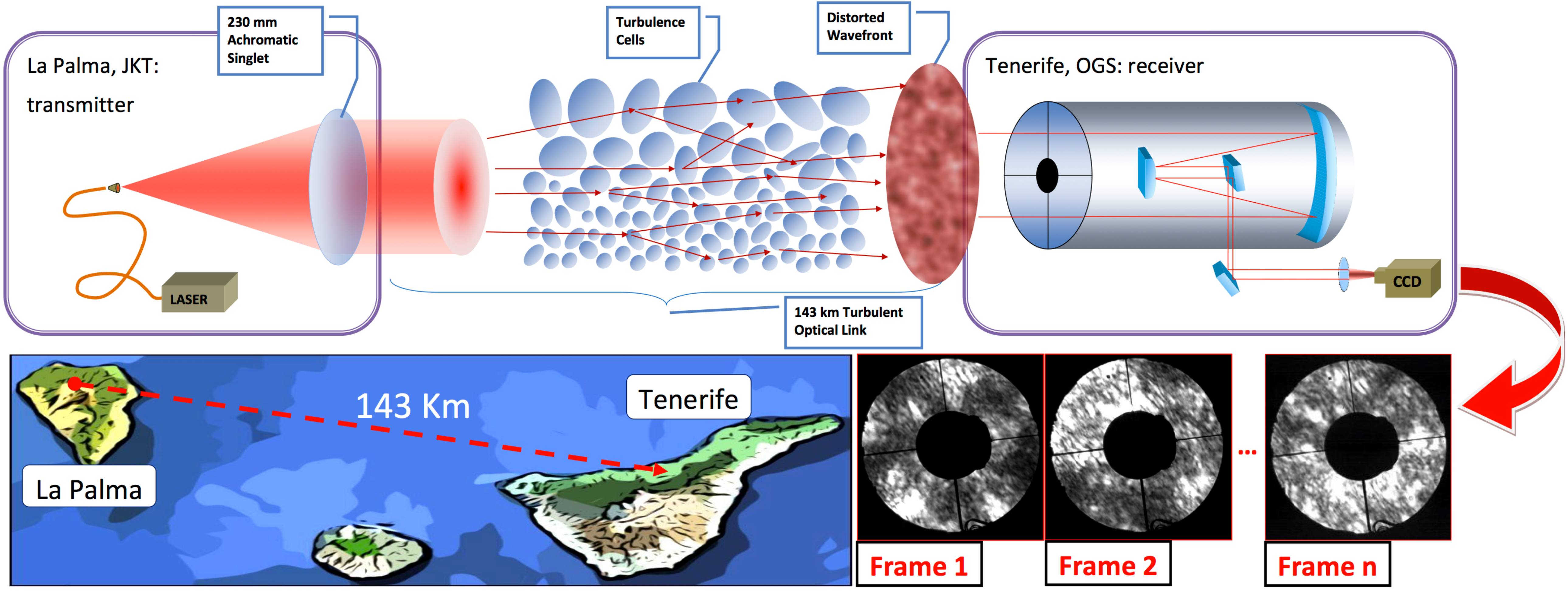}
\caption{Experimental setup.
 At the transmitter side in La Palma, a  $\lambda=810 \, nm$ laser beam is collimated with a $230 \, mm$ achromatic singlet, explicitly realized to limit geometrical distortions, and then sent through a $143 \, km$ free space optical channel. At the receiver side, at the OGS observatory in Tenerife, the pupil of the $1016 \, mm$ Ritchey-Chr\'etien telescope  is illuminated by the distorted wavefront
and imaged on a high resolution CCD camera. }\label{atmolayer}
\end{figure*}

Coherent light propagation along scattering media provide examples of randomness.
More specifically, when coherent light crosses a random scattering medium, the complex field undergoes subsequent diffusion
process either described within a random walk or L\'evy flight \cite{barthelemy2008levy} model, giving rise to a random distribution of the intensity as consequence of the interference effects. For instance, static speckle patterns obtained  by passing {a} laser beams through volumetric scatterers \cite{Marron1986,Horstmeyer2012} 
have been already exploited for the purpose of random number generation and as key element of physical un-clonable functions \cite{Pappu2002}.
However, these approaches are based on still scattering medium and cannot be used for real time random number generation.

In this Letter, we propose and realize a TRNG which converts the chaotic and rapidly varying dynamic of atmospheric turbulence in random bits.
For a proof of principle of the method, {we send a coherent laser beam} through a free space channel of $143 \, km$ used in the last years for experiments in Quantum Communications \cite{ursi07nap, sche10pnas, ma12nat, capr12prl}, which give rise to an unprecedented
long link devoted to  this purpose.
The chaotic fingerprint of the turbulence is transferred to the laser beam propagation giving rise to an extended speckle pattern
which was acquired by using a camera. An optimal coding theory is then applied on the single realizations of the beam to extract random bits from the images. The source of entropy is {atmospheric turbulence}, a macroscopic physical process which cannot be controlled or influenced by an attacker. In addition, the simplicity of the method does not require the use of front-end devices which are usually prone to instabilities and failures.

The unfeasibility
{of analytical and numerical solutions} of the Navier-Stokes equations for the {atmospheric} turbulent flow, led Kolmogorov to develop his well known statistical model \cite{kolmogorov1941local}, which parameterizes the repartition of {kinetic} energy as the interaction of decreasing size \textit{eddies}. The dynamic of the eddies is mainly ruled by temperature variations, which cause fluctuations in the air refractive index $n$. Turbulent atmosphere is then comparable, when a laser beam is sent across it, to a dynamic volumetric scatterer: after the propagation, speckle patterns are observed since the field is subjected to phase delay and amplitude fluctuations, induced by the inhomogeneities of the atmosphere
\cite{Fante1975, fante1980electromagnetic, andrews2005laser}.  It is worth to stress that the theoretical framework of the atmospheric turbulence, allow us to assume that we are dealing with a system randomly evolving. Indeed, the phase space trajectories which are strongly dependent on the initial conditions, start to be separated in time scales which are much smaller than the time scale we sample the system.

For the purpose of sampling the highly entropic content of optical turbulence, we established a free space optical (FSO) link $143 \, km$ long by a sending $\lambda=810 \, nm$ laser beam between the \textit{Jacobus Kaptein Telescope} (JKT) in the Island of La Palma, to the ESA \textit{Optical Ground Station} (OGS) in the Island of Tenerife, see Figure \ref{atmolayer} for details.  At the receiver telescope, the effects 
of beam propagation in strong turbulence are present: wandering, spreading, wave-front breaking and scintillation \cite{Fante1975}. 
The motion of eddies larger {than} the beam cross section, bend it and causes a random walk of the beam center on the receiver plane. Whereas, small scale inhomogeneities diffract and refract different parts of the beam which then {constructively and destructively interfere} giving {rise to} a speckle pattern on the telescope pupil. Both the previous factors spread the beam beyond the inherent geometrical limit.  Furthermore, {it is possible to observe scintillation, namely} fluctuations in the irradiance of the signal.
Since the eddies are continuously  moving according the chaotic flow of the atmosphere, the distribution of the scintillation peaks in the receiver plane evolves randomly. So, for the purpose of random number generation, we acquired images with an exposure time shorter than the characteristic time of fluctuations in order to not {average} out the dynamic of the process.

For the implementation of the method, the CCD relevant pixels are labelled sequentially with an index $s$, $s \in \{1, \dots, N\}$ and the $n_f$ speckle centroids of the frame $f$ are elaborated (for details on the centroid extraction see Supplementary Informations). By considering then the pixels where a centroid fall in, an ordered sequence
$S_f=\{s_1 ,s_2 ,\dots,s_{n_f} \}$
with $s_1<s_2<\dots <s_{n_f}$,
can be formed. In this way the pixel grid can be regarded as the classical collection of urns where the turbulence randomly throw balls (the centroids) in.  Because of  the random nature of the process, the centroids visit every part of the grid with the same probability:
a given frame $f$ ``freezes"  one $S_f$ out of the
\begin{equation}
T_f=\frac{N!}{(N-n_f)!n_f!}
\end{equation}
possible and equally likely sequences of $n_f$ centroids. Among all of them,
 a given $S_f$ can be univocally identified with its lexicographic index $I(S_f)$
\begin{equation}\label{main}
I(S_f)=\sum^{n_f}_{k=1}
\binom{N-s_{k}}{n_f-k+1}
\end{equation}
with $0\leq I(S_f) \leq T_f-1$. Basically, (\ref{main})  enumerates all the possible arrangements which
\textit{succeed}
a given centroids configuration (see Supplementary Informations)
and the TRNG distillates randomness by  realizing the correspondence $S_f\Longleftrightarrow I(S_f)$. Indeed, as an uniform RNG is supposed to yield numbers \textit{identically and independently distributed} (i.i.d.) in a range  $\left[X,Y\right]$, as this method generates a random integer in the range $\left[0,T_f-1\right]$.

\begin{figure}[!h]
\centering
\includegraphics[scale=0.16]{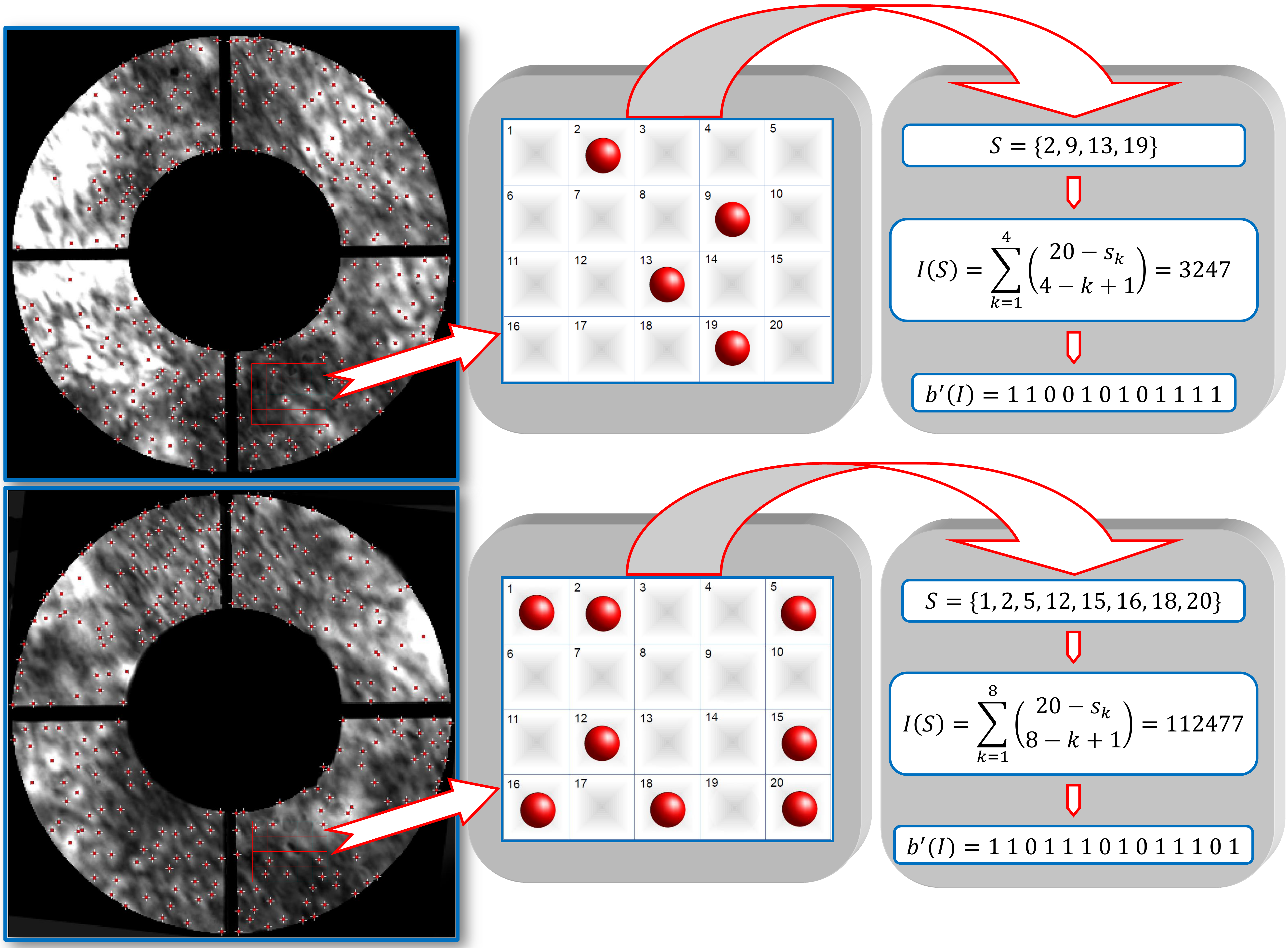}
\caption{In Figure two  sample frames are reported, with the centroids of the brightest speckles evaluated.
It is worth to stress that for illustrative purposes the image has been simplified: in the real implementation centroids are evaluated on different intensity levels and every cell corresponds to a pixel. Let's consider 20 urns (the pixels) and 4 balls (the centroids) as in top figure.
The total number of combinations is $T=\binom{20}{4}=4845$ with $L=\lfloor \log_2T\rfloor=12$. The ball positions are defined by the sequence
$S\equiv\{s_1,s_2,s_3,s_4\}=\{2,9,13,19\}$ that corresponds to the lexicographic index $I(S)=3247$. Since $I(S)<2^L$,
12 bits, i.e. the binary expansion of $I(S)$ ``110010101111", can be extracted from $S$.
When $I(S)\geq 2^L=4096$ less bits can be extracted (see Supplementary Informations). A similar procedure for the bottom figure with 8 balls in 20 urns
allow to extract the sequence $b'(I)=11011101011101$.
}\label{urns}
\end{figure}

To be conveniently handled, a binary representation  $b_{I_f}$ of the random integers $I(S_f)$ must be given.
The simpler choice is to transform the integer $I(S_f)$ in binary base, obtaining a sequence  with $L_{T_f}=\lfloor\log_2{T_f}\rfloor$ bits.
However, only if  $T_f \, \text{mod} \, 2^i =0$ for $i\in \mathfrak{N}$, every frame $f$ would theoretically provide strings $L_{T_f}$ bits long.
In general this is not the case and hence, all the frames with $\log_2{I(S_f)} \geq  L_{T_f}$ should be accordingly discarded to avoid the so-called \textit{modulo bias}. This issue, which clearly limits the rate of generation,
can be solved by adopting an encoding function $E: b_{I_f}\rightarrow E\left[b_{I_f}\right]\equiv b_{I_f}^\prime$ developed by P. Elias \cite{elia72anm} which re-maps strings longer than  $L_{T_f}$ in sets of shorter sub-string with the same probability of appearance
(see Supplementary Informations).
This approach is optimal: the positions of $n_f$ centroids in $N$ pixels can be seen as a biased sequence of $N$ bits,
with $n_f$ ones and $N-n_f$ zeros. The content of randomness of this biased sequence is $h_2(q)=-q\log_2q-(1-q)\log_2(1-q)$ with
$q=\frac{n_f}{N}$. By the Elias method it is possible to unbias the sequence in an optimal way:
it can be shown that the efficiency
$\eta=\frac{\langle  L_{b^\prime}\rangle}{N}$,
the ratio between the average length of {$b'_{I_f}$ and $N$}, reaches the binary entropy $h_2(q)$ in the limit of large
$N$, $\lim_{N \to\infty}\eta=h_2(q)$.
In this way it has been possible to preserve the i.i.d. hypothesis for the set $\left[0,1\right]$ maximizing the rate of the extraction.

It is worth noting that the combinatorial
approach here presented to generate random number is completely general and can be applied in many situations in which
several spots are randomly distributed in a frame, without the requirement of having an average number of spots that is
the half of the possible spot positions.

\begin{table}[!t]
\centering
\includegraphics[scale=0.8]{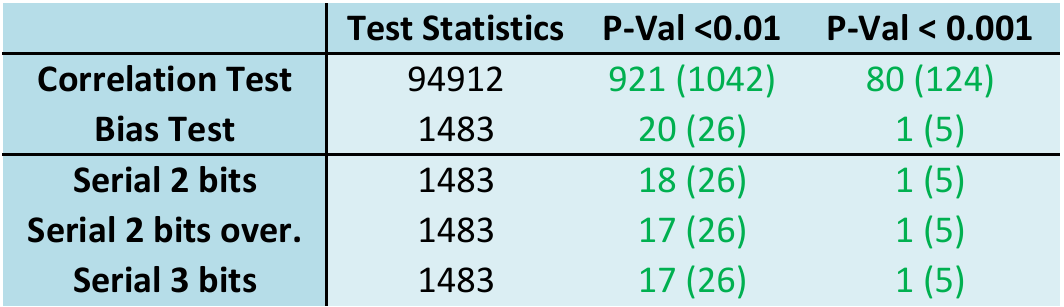}
\caption{In table, for every test (first column) the overall number of tests statistics (second column)  obtained from videos recorded in different conditions are reported. The number of failures are listed in the third and fourth columns. These numbers can  be compared with the theoretical number of failures (inside the paranthesis) which are expected when the i.i.d. hypothesis hold true. As it can be seen for all the tests the failures are inside the limits both for the 99\% and 99.9\% confidence levels.}\label{smalltable}
\end{table}
\begin{figure}[!t]
\centering
\includegraphics[scale=0.22]{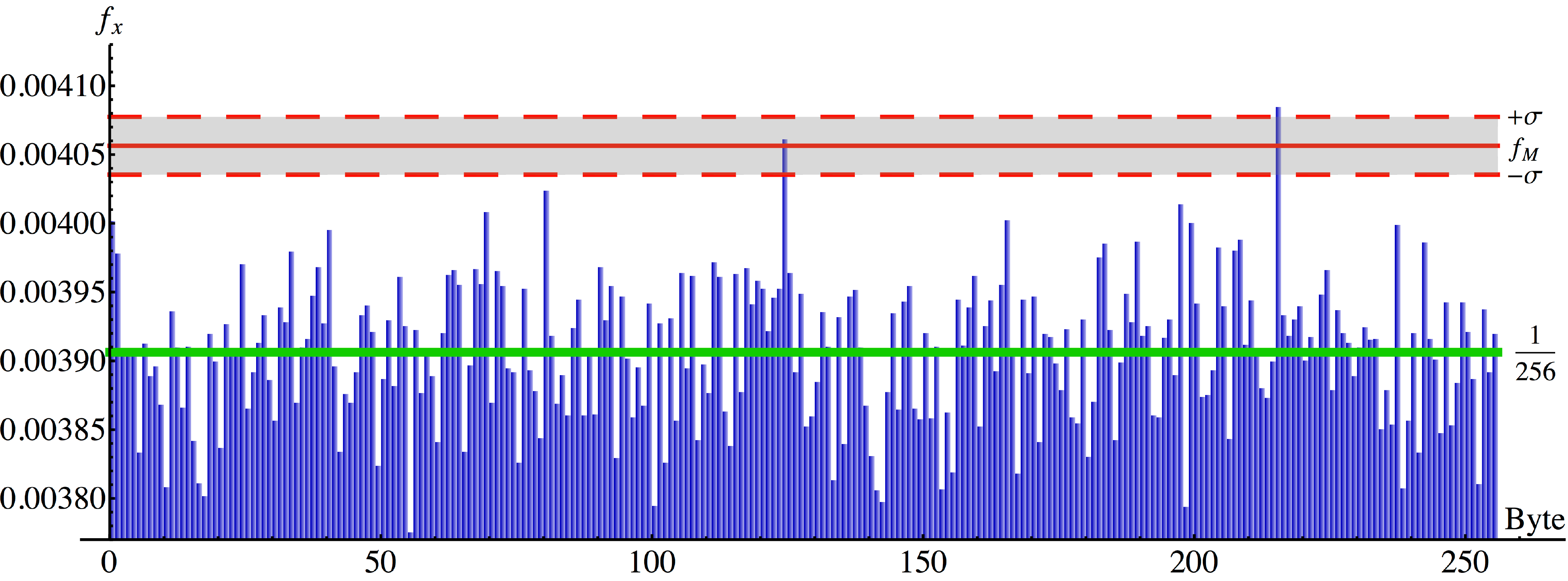}
\caption{The histogram shows a detail of the byte occurrence relative frequencies obtained from $1.4\cdot 10^6$ bytes
(corresponding to 671 frames). The evaluated chi-square statistic on the frequency yields a $\mathcal{P}\text{-val}=0.77$. {One can notice how the frequencies randomly distribute at the sides of the expected mean value (green line).
Furthermore, the maximal byte frequency (corresponding to the byte 215) it is fully compatible with its expected value (red solid line)
 and the  $\pm\sigma$ limits (red dashed lines).} }\label{histobig}
\end{figure}
By implementing this technique with different configurations of masks and centroids,  we were able to reach a maximum average rate of
$17\  {{kbit}}/{{frame}}$ (with a grid of 891000 urns and an average of 1600 centroids per frame). It is worth to stress that, for the present proof of principle, the distillation of random bits has been done offline so, theoretically, having used a frame rate of
$24\ {frame}/{s}$ this method could provide a rate of $400 \, {kbit}/{s}$ using a similar camera and further increasing such
rate by using a larger sensor.
\begin{table*}[!t]
\centering
\includegraphics[width=\textwidth,height=5cm]{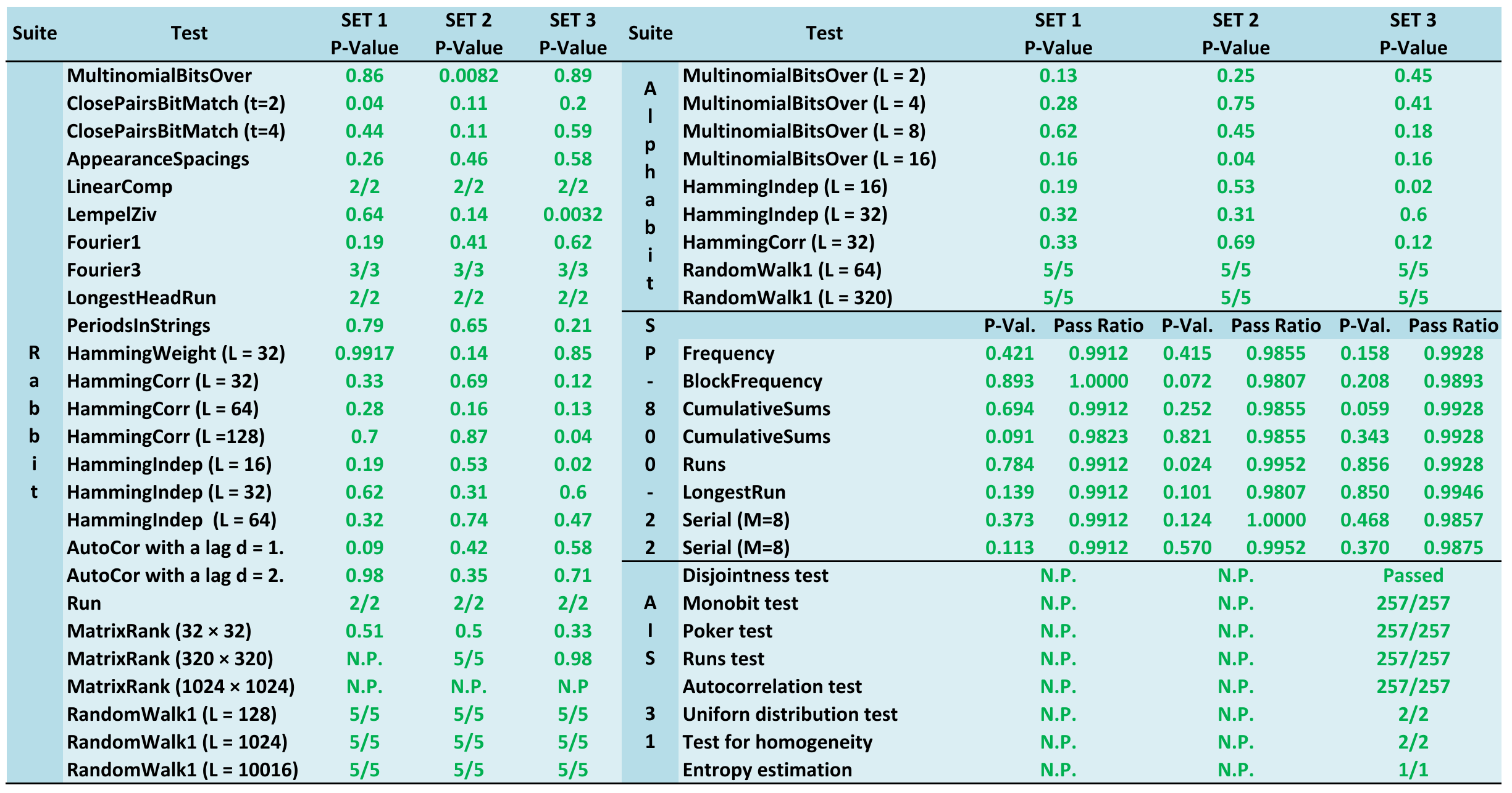}
\caption{\footnotesize{In table, the results of selected tests of batteries, particularly effective in detecting defects in TRNG are summarized. The \textit{Alphabit} and \textit{Rabbit} batteries belong to the \textsc{TESTU01}: critical results are if $\mathcal{P}\text{-val} \leq 10^{-3}$ or $\mathcal{P}\text{-val} \geq 0.990$. The NIST SP-800-22 suite has passing ratio critical values for the three sets equal to 0.95575, 0.96618 and 0.97674 respectively. The test on the distribution of  p-values must be $\mathcal{P}\text{-val} \geq 10^{-5}$. The AIS31 suite could be applied only on the larger set of bits: as it can be seen all the 263 tests of this suite were passed. }}\label{grid}
\end{table*}

For assessing the randomness of a TRNG, in addition to a sound knowledge of the physical process employed, it is necessary to apply statistical tests in order to exclude the presence of defects caused by a faulty hardware. The output of a test on a bit string is another random variable with a given distribution of probability, the so-called \textit{test statistic}.
Hence, we computed the $\mathcal{P} \text{-value}$, namely the probability of getting
 an equal or worse test statistic, holding true the i.i.d. hypothesis.
If the $\mathcal{P} \text{-values}$ are smaller than some a priori defined critical values the tests are considered failed: these limits are
are usually chosen as $\mathcal{P} \text{-value} <0.01$ and $\mathcal{P} \text{-value} <0.001$,
corresponding to a confidence level of 99\% and 99.9\% respectively.
Otherwise, whenever one obtains $\mathcal{P} \text{-values}$ equal or greater than these limits,  the i.i.d. hypothesis for the tested string is assessed.

As first result of the statistical analysis, we present the outcomes of two tests, the \textit{frequency} and the \textit{autocorrelation} test respectively \cite{menezes2010handbook}. The first test checks whether the fraction of 0s and 1s departs from the expected value of $\nicefrac{1}{2}$ over the acceptable statistical limits. The second test evaluates whether the bit values depend on the neighbouring bits.
The output of both the tests (the serial autocorrelation with bit lag from 1 to 64) are test-statistics normally distributed and the analysis results are reported in Table \ref{smalltable}: the number of frames for which the  i.i.d. hypothesis does not hold
with a confidence of 99\% and 99.9\% (corresponding to  $\pm2.58\sigma$ and $\pm3.29\sigma$ respectively) are inside the critical limits of statistical fluctuations.

The main consequence of major defects absence at single bits level, is an even repartition of the Hamming weights which allows to pass also the so-called \textit{serial tests} for the uniform distribution of many bits words. Applied on 2-bits, 2-bits overlapped and three 3-bits words the tests were all passed, \ref{smalltable}.

The suitability of the method for random number generation depends on the statistical properties of the atmospheric turbulence over the time, in other words the stationarity and ergodicity of the physical process employed. It has been then fundamental to check the i.i.d. hypothesis for the numbers obtained by joining the bits belonging to frames of the same videos. A visual evidence that an overall uniformity is preserved during the whole acquisition time, it is given in Figure \ref{histobig} where the distribution of $1.4\cdot 10^6$ bytes obtained from a 671 frames video sample is plotted. If the bytes were used for cryptographic purposes, it is meaningful to consider the binary \textit{min-entropy}
$h_{min}=\max_i[- \log_2(p_i)]$ where $p_i$ is the measured appearance probability of the byte $i \in \left[0,255\right]$. A value of
$h'_{min}=7.936$
bits per byte has been measured and this can be compared with the expected min-entropy for a sample of that size, that is
$H_{min}=7.946 \pm 0.007$. The experimental obtained value is thus in agreement with the expected value from theoretical prediction
on uniform distribution, assessing an eavesdropper has no advantage with respect to random guessing 
(see Supplementary Informations for a derivation of the expected min-entropy $H_{min}$).

The get a confirmation of the i.i.d. hypothesis for the whole sets of bits, the numbers were thoroughly analysed with three state-of-the-art  batteries of tests whose results are reported in Table \ref{grid}. At present time, the \textsc{TEST-U01}  \cite{Simard2009} is a most stringent and comprehensive suite of tests; among all, we chose a pair sub-batteries, \textit{Rabbit} and \textit{Alphabit} respectively, specifically designed to tests TRNGs. Note that, other batteries designed for algorithmic generators do not include tests sensitive to the typical TRNGs defects, such as correlations and bias. As it can be seen all the results were outside the limits of $\mathcal{P}\text{-val} \leq 10^{-3}$ or $\mathcal{P}\text{-val} \geq 0.990$.
The \textsc{SP-800-22} \cite{Rukhin2001} is developed by the \textit{NIST} and it represents a common standard in RNG evaluation. For this suite, the files were partitioned in sub-strings $20000$ bits long and the tests suitable for this bit-length were applied with the recommended parameters. Also in this case we registered successful results, being both the ratio between the sub-strings with $\mathcal{P}\text{-val} \geq 10^{-2}$
and the total number of strings and the second-order test on $\mathcal{P}\text{-val}$ distribution, over the critical
limits\footnote{Passing ratios depends, time to time, on the number of strings analysed, see Table \ref{main}. For the goodness-of-fit test on the p-value distribution the limit is $\mathcal{P}\text{-val} \leq 10^{-5}$.}.
Eventually, on the largest file obtained, we successfully applied also the \textsc{AIS-31} \cite{Killmann2011} suite developed by the German BSI.
The \textsc{AIS-31} offers three sub-batteries of increasing difficulty which are intended to be applied \textit{on-line}, that is to monitor the output of TRNG in order to detect failures and deviation from randomness: according to which level is passed, a TRNG can be considered preliminary suitable for different purposes (T0 pre-requisite level, T1 level for TRNGs used in connection with PRNG, T2 level for stand-alone TRNGs).

{As conclusion from the more stringent tests, the i.i.d. hypothesis resulted confirmed and strengthened.
We would like to point out that our method do not rely on sensitive and hardly detectable processes which require extremely tuned hardware: indeed unavoidable hardware non-idealities (such as drifts in voltage threshold levels, asymmetric beam-splitters, photo-detector afterpulses, etc.) can induce bits dependencies and bias. In this case, the generator will fail the tests and some algorithmic randomness extractor
is required to obtain true random strings. The fact that here the randomness is assessed without the need of any post-processing technique, is of particular significance to demonstrate the effectiveness of the present method.}

\begin{center}
\noindent\textbf{\large Supplementary Information}
\end{center}

\noindent\textbf{Image processing.} To extract the randomness from the frames of the videos,  typical algorithms for image analysis which allows to compute several so-called digital \textit{moments} were employed. More precisely, given $E$ the number of bits used by the acquisition software to encode the intensity (colour) levels of monochromatic light on the active area $m \cdot n$ of the sensor, we can consider the recorded image as a two variables function $I(x,y)$ where $x \in \{0, \dots, m\}$, $y \in \{0, \dots, n\}$ and $I(x,y) \in \{0, \dots, 2^E\}$. The $(j,k)^{th}$ moment of an image is then defined as

\begin{equation}
M^{jk}=\sum_{x=1}^m\sum_{y=1}^n I(x,y) x^j y^k \qquad .
\end{equation}

The \textit{center of gravity} $C$ (also known as centroid) of an image is then located at position $(\widehat{x},\widehat{y})$ where the coordinates are accordingly given by
\beq
\widehat{x}=\frac{M^{10}}{M^{00}}\,,\qquad
\widehat{y}=\frac{M^{01}}{M^{00}}
\eeq
We applied then a technique for instance used in Biology to count the number of cells in biological samples. Indeed in images composed by distinguishable components (as coloured cells on a uniform background), it is possible to \textit{locally} calculate the centroids $C_i$ of those components, by binarizing the intensity level, i.e. by setting $E=1$, and then evaluating the moments on the  closed subsets $S_i=\{(x,y)|I(x,y) = 1\}$, that is
\begin{equation}
M_{jk}(S_i)=\sum_{(x,y) \in S_i} I(x,y) x^j y^k
\end{equation}
where the index $i$ runs on the different elements of the image.

To extract more randomness from the geometrical pool of entropy, the intensity profile of the frames has been partitioned into eight different sub-levels.
We treated separately every different intensity level, $L$,  as a source of \textit{spots}; more specifically then we generated sets $S_{L,i}$ out of the $L \in \{1, \dots, 8\}$ levels. 
For a given $L$ and a spot $i$ the coordinates of a centroids are then
\beq
\widehat{x}_{L,i}=\frac{1}{A_{i,L}}\sum_{x \in S_{i,L}} x \,\qquad
\widehat{y}_{L,i}=\frac{1}{A_{i,L}}\sum_{y \in S_{i,L}} y
\eeq
where $A_{i,L}$ simply the area of the spot, that is the total number of pixels which compose that spot.
In order to remove edge effects due to the shape irregularities of the pupil, pixels close to irregular edges were removed.

{\noindent \textbf{Bit extraction.}  Let's now demonstrate formula 2 of the main text.
We need to enumerate the combination of $n_f$ balls contained in $N$ urns. The positions
of the ball are identified with the integers $s_1<s_2<\cdots<s_{n_f}$. The number of possible combinations is $T_f=\binom{N}{n_f}$.
Let's first calculate the number of combinations that
precede the given combination. This can be obtained by summing all the possible combinations in which the first ball falls
in the positions $s'_1$ with $s'_1<s_1$, namely $\sum_{m=1}^{s_1-1}\binom{N-m}{n_f-1}$,
plus all the combination in which the first ball is in $s_1$ and the second ball is in $s'_2$ with $s_1<s'_2<s_2$,
namely $\sum_{m=s_1+1}^{s_2-1}\binom{N-m-1}{n_f-2}$,
plus all the combination in which the first ball is in $s_1$, the second in $s_2$ and the third ball is in $s'_3$ with $s_2<s'_3<s_3$ and so on.
This number is given by
\beq
p(S_f)=\sum^{n_f-1}_{k=0}
\sum^{s_{k+1}-1}_{m=s_{k}+1}
\binom{ N-m }{n_f-k-1}
\eeq
where we defined $s_0=0$. From $\sum _{k=0}^n \binom{k}{j}=
\binom{ n+1}{ j+1}$, it can be shown that
$\sum _{m=s_k+1}^{s_{k+1}-1} \binom{N-m}{n_f-k-1}=\binom{N-n_k}{n_f-k}-\binom{N-n_{k+1}+1}{n_f-k}$
so that $p(S_f)=\binom{N}{n_f}-\sum _{k=1}^{n_f}\binom{N-s_k}{n_f-k+1}-1$.
The number of combination that succeed $S_f$ can be easily computed by
\beq
I(S_f)=\binom{N}{n_f}-1-p(S_f)=\sum _{k=1}^{n_f}\binom{N-s_k}{n_f-k+1}
\eeq
where $0\leq I(S_f)<T_f$.}
The number $T_f-1$ represents then the upper bound to the uniform distribution of arrangement indexes which can be obtained by all the possible arrangements of $n_f$ centroids: the largest index, that is $I(S_f)=T_f-1$, is when all the centroids occupy the first  urns of the grid.

To convert the integer $I(S_f)$, uniformly distributed in the interval $[0, T_f-1]$, into an unbiased sequence of bits, we adopted
an optimal binary encoding strategy introduced by P. Elias  \cite{elia72anm}.
Let's consider  the binary expansion of $T_f$
\begin{equation}
T_f=\alpha_L \cdot 2^L+\alpha_{L-1} \cdot 2^{L-1}+\dots+\alpha_0 \cdot 2^0
\end{equation}
where $L=\lfloor\log_2T_f\rfloor$, $\alpha_L=1$ and $\alpha_k = 0,1$ with $0 \leq k< L-1 $.
Random bit strings are associated to $I(S_f)$ according to the following rule:
find the greatest $m$ such that
\beq
I(S_f)<\sum_{k=m}^L\alpha_k 2^k
\eeq
and extract the first $m$ bits of the binary expansion of $I(S_f)$. By this rule,
when $I(S_f)< 2^L$, $L$ bits can be extracted;
when $2^L\leq I(S_f)< 2^L+\alpha_{L-1}2^{L-1}$, $L-1$ bits can be extracted and so on; when $I(S_f)=T_f-1$ and $\alpha_0=1$
 (namely when $m=0$) no string is assigned. It can be easily checked that this method produces unbiased sequences of bits
from integers uniformly distributed in the interval $[0, T_f-1]$.

\noindent\textbf{Min-entropy estimation.} In this section we show how to estimate the expected min-entropy.
In a sample with $L$ bytes,
the single byte occurrence $\ell_i$ ($i=1,\cdots 256$)
are random variables distributed according the Poisson distribution with mean $\lambda=\frac{L}{256}$.
In order to estimate the expected min-entropy we need the distribution of the maximum of the occurrences and we can proceed as follow.
Given a sample of $n$ random variables $X_1, X_2, \dots, X_n$ whose cumulative distribution function (CDF) is $D(x)$ and the probability density function (PDF) is $F(x)$, they can be re-ordered as {{$X_{\pi(1)}\leq X_{\pi(2)}\leq \dots \leq X_{\pi(n)}$}}:
the $X_{\pi(k)}$ is called statistic of order $k$, such that $\min \left\{ X_1, X_2, \dots, X_n \right\}=X_{\pi(1)}$ and 
$\max \left\{X_1, X_2, \dots, X_n \right\}=X_{\pi(n)}$.
In order to derive the distribution function of an order $k$ statistic, given $h$ the number of $X_i\leq x$, one can note that
\begin{equation}
\begin{aligned}
  D_{k}(x) &= P(X_{\pi(k)} \leq x)= P(h \geq k) = \sum_{i=k}^n P(h=i)
  \\
  &= \sum_{i=k}^n \binom{n}{i}[D(x)]^k[1-D(x)]^{n-k}
\end{aligned}
\end{equation}
Working with integer random variables the PDF is then obtained by
\begin{equation}
F_{k}(x) = D_{k}(x)-D_{k}(x-1)
\end{equation}
Being interested in the byte frequencies maximal values, that is $k=n$, the previous equation becomes
\begin{equation}
F_{n}(x)
 =  [D(x)]^n-[D(x-1)]^n
\end{equation}
In a sample with size $L$, the distribution of the maximum $\ell_M$ of the single byte occurrence $\ell_i$ can be computed
by using the previous equation with $D(x)=e^{-\lambda}\sum_{j=0}^x\frac{\lambda}{j!}$, $\lambda=\frac{L}{256}$ and n=256:
\begin{equation}
\begin{aligned}
\Pi(\ell_M)
 &= (e^{-\lambda}\sum_{j=0}^{\ell_M}\frac{\lambda}{j!})^n-(e^{-\lambda}\sum_{j=0}^{\ell_M-1}\frac{\lambda}{j!})^n \\
&=  \left(\frac{\Gamma (\ell_M+1,\lambda )}{\ell_M!}\right)^n-\left(\frac{\Gamma (\ell_M,\lambda )}{\Gamma (\ell_M)}\right)^n
\end{aligned}
\end{equation}
The expected value and variance of the maximum of the $\ell_i$'s, are then easily evaluated by applying the definitions
$\langle \ell_M\rangle =\sum_{x=0}^{\infty}x\Pi(x)$ and
 and  $\sigma^2=\langle \ell^2_M\rangle-\langle \ell_M\rangle^2$
  respectively. With a sample size of $L=1399852$ bytes
 and $n=256$, the theoretical reference values are then evaluated to be
 $\langle \ell_M\rangle =5678.4 \pm 29.4$
  counts with
 corresponding expected relative frequency
 $\langle f_{M}\rangle=\frac{\langle \ell_M\rangle}{L}=(4.056\pm0.021)\cdot 10^{-3}$.
  This value corresponds to a theoretical min-entropy of $H_{min}=-\log_2f_M=7.946 \pm 0.007$ bits per byte.
If the obtained experimental min-entropy is compatible with the predicted theoretical value, the sample
can be considered as uniformly distributed. 

\noindent\textbf{\it Acknowledgements} 
The authors wish to warmly thank for the help provided by Z. Sodnik of the European Space Agency and by C. Barbieri and S. Ortolani of University of Padova as well as by the Instituto de Astrofisica de Canarias (IAC), and in particular F. Sanchez-Martinez, A. Alonso, C. Warden and J.-C. Perez Arencibia, and by the Isaac Newton Group of Telescopes (ING), and in particular M. Balcells, C. Benn, J. Rey, A. Chopping, and M. Abreu. 

This work has been carried out within the Strategic-Research-Project {\bf QUINTET} of DEI-University of Padova and the Strategic-Research- Project {\bf QUANTUMFUTURE} of the University of Padova.

\end{document}